# *Proof of the bulk-edge correspondence through a link between topological photonics and fluctuation electrodynamics*

*Mário G. Silveirinha*[*]

[1] *University of Lisbon–Instituto Superior Técnico and Instituto de Telecomunicações, Avenida Rovisco Pais, 1, 1049-001 Lisboa, Portugal, mario.silveirinha@co.it.pt*

**Abstract**

The bulk-edge correspondence links the Chern-topological numbers with the net number of unidirectional states supported at an interface of the relevant materials. This fundamental principle is perhaps the most consequential result of topological photonics, as it determines the precise physical manifestations of nontrivial topological features. Even though the bulk-edge correspondence has been extensively discussed and used in the literature, it seems that in the general photonic case with dispersive materials it has no solid mathematical foundation and is essentially a conjecture. Here, I present a rigorous demonstration of this fundamental principle by showing that the thermal fluctuation-induced light-angular momentum spectral density in a closed cavity can be expressed in terms of the photonic gap Chern number, as well as in terms of the net number of unidirectional edge states. In particular, I highlight the rather fundamental connections between topological numbers in Chern-type photonic insulators and the fluctuation-induced light-momentum.

---

[*] To whom correspondence should be addressed: E-mail: *mario.silveirinha@co.it.pt*

## I. Introduction

Topological matter and topological systems can have rather exotic properties and very unusual physics. In particular, systems with a broken time reversal symmetry (Chern-type insulators) are usually characterized by a topological index known as the Chern number [1-9]. In the fermionic case, the Chern number determines the quantized Hall conductivity of a 2D electron gas in the zero-temperature limit [10-14].

One of the most significant and far-reaching results in topological photonics is the so-called "bulk-edge correspondence" [3, 5, 15]. This fundamental principle links the Chern invariants of two photonic insulators with the net number of unidirectional edge states supported by an interface of the two materials. A recent work reported a proof of the bulk-edge correspondence for a bosonic Bogoliubov-de Gennes Hamiltonian over a tight-binding Hilbert space [16]. However, notwithstanding the bulk-edge correspondence has been extensively discussed and used in the recent literature, it seems that so far, for photonic crystals formed by dispersive materials, it has no solid mathematical foundation and is mainly a conjecture (see a discussion in [8]). Indeed, the arguments in favor of the bulk-edge correspondence are mainly heuristic, e.g., that a continuous transformation of one photonic "mirror" into another topologically distinct mirror requires closing the band gap, and that thereby a material interface must support edge-states. Alternatively, they rely on analogies with the electronic case, for which there are compelling physical reasons to avow that the bulk-edge correspondence holds [10, 12], and mathematical derivations for some two-dimensional systems [17-20]. However, given the different nature of fermionic and bosonic systems, the extrapolation of the condensed-matter arguments to optics is at least questionable. Furthermore, the analogies between electronics and optics are typically valid in a limited quasi-momentum range, e.g., in the framework of some tight-binding approximation limited to some section of the Brillouin zone, whereas the topological invariants are determined by the global properties of the Hamiltonian. Some photonic systems may be mapped onto either a single-particle fermionic system (see Refs. [21, 22]) or possibly onto a Bogoliubov-de Gennes lattice model [16] in the entire Brillouin zone. The use of the bulk-edge correspondence is evidently justified in such cases, but these are arguably the exceptions rather than the rule.

Different from most studies of topological photonics, the configuration under analysis here consists of topological material enclosed in a cavity (Fig. 1a). In closed systems the edge states are forced to circulate around the cavity walls and this may lead to novel physical effects. In the recent work [23], I showed that the thermal fluctuation-induced light angular momentum density per unit of area is precisely quantized in the photonic-insulator cavity, and that its "quantum" is determined by the net number of unidirectional edge-states circulating around the cavity. This rather universal property holds even when the system has no topological classification. In particular, in nonreciprocal platforms the thermal equilibrium condition is compatible with a persistent energy circulation in closed orbits [23-26].

The quantization of the fluctuation-induced angular momentum can be explained in a simple and intuitive manner. Specifically, it may be observed that in the band-gaps of the bulk region the allowed photonic states are necessarily edge waves; thus the topological cavity may be regarded as a circular multi-mode transmission line (Fig. 1b).

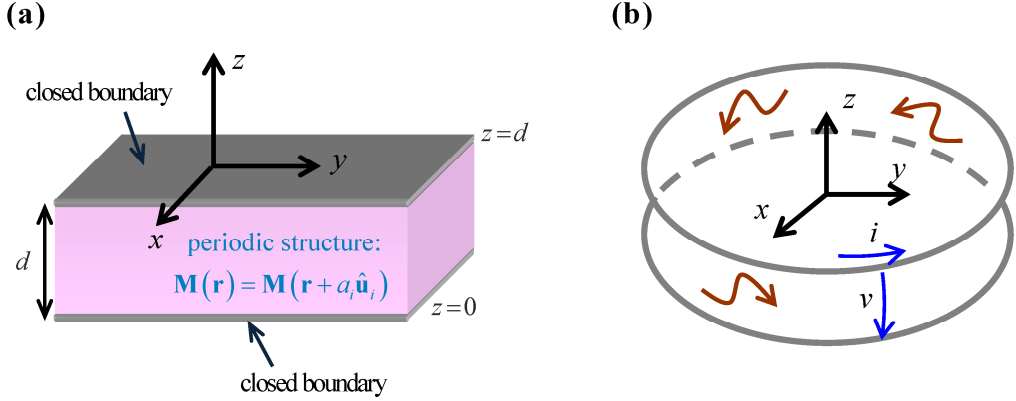

Fig. 1 (a) Representative geometry of the system under-study. The bulk region is periodic along the *x* and *y* directions ($\mathbf{M}(\mathbf{r}) = \mathbf{M}(\mathbf{r} + a_i \hat{\mathbf{u}}_i)$, with $a_1, a_2$ the spatial periods along *x* and *y*). The structure cross-section has area $A_{tot} = L_1 \times L_2$ and generally encompasses many elementary cells of the associated photonic crystal. The system is closed along *z* so that the energy is forced to flow along directions parallel to the *xoy* plane. In this study there are two cases of interest: *(i)* the fields satisfy opaque-type (e.g., perfectly electric conducting) boundary conditions at the lateral-walls (not shown) or *(ii)* the fields satisfy periodic boundary conditions at the lateral walls. (b) For opaque-type lateral walls, the photonic insulator cavity is equivalent to a circular one-dimensional transmission line.

As it is well-known, thermal and quantum fluctuations can induce energy flows in a transmission line terminated with resistive loads [28]. For a system in thermal equilibrium, the power transported by a certain travelling wave is $p_\omega d\omega$, with $p_\omega = \dfrac{\mathcal{E}_{T,\omega}}{2\pi}$ the power spectral density, $\mathcal{E}_{T,\omega}$ the mean energy of a quantum-harmonic oscillator at temperature $T$ ($\mathcal{E}_{T,\omega} \approx k_B T$ when $\hbar\omega \ll k_B T$) and $d\omega$ is the relevant bandwidth [28]. The light angular momentum can be found by integrating the Abraham angular momentum density, $\mathbf{r} \times \mathbf{S}/c^2$ with $\mathbf{S}$ the Poynting vector, over the cavity volume. Evidently, for a *circular* line the fluctuation-induced light angular momentum per mode ($\mathcal{L}_{T,\omega}^{\text{mode}} d\omega$) is determined by the spectral density $\mathcal{L}_{T,\omega}^{\mathrm{mode}} = \pm \frac{1}{c^2} p_\omega 2 A_{tot} = \pm \frac{1}{c^2} \frac{\mathcal{E}_{T,\omega}}{\pi} A_{tot}$. The + (–) sign is chosen for waves travelling in the anti-clockwise (clockwise) direction and $A_{tot}$ is the cavity cross-sectional area. In the reciprocal case, the number of travelling waves propagating in the anti-clockwise and clockwise directions is identical; hence the net fluctuation-induced angular momentum vanishes: $\frac{\mathcal{L}_{T,\omega}}{A_{tot}} = 0$. In contrast, in a nonreciprocal system the number of edge states circulating in the two directions may be different. Thereby, the net angular momentum density is given by $\frac{\mathcal{L}_{T,\omega}}{A_{tot}} = N \frac{1}{c^2} \frac{\mathcal{E}_{T,\omega}}{\pi}$ and is fully determined by the net number $N$ of edge states circulating in the anti-clockwise direction, which is the result of [23]. In Ref. [23], the "bulk-edge correspondence" was used to link the angular momentum "quantum" ($N$) with the gap Chern number.

Here, I follow precisely the inverse path. It is demonstrated –never making use of the bulk-edge correspondence– that there is an intimate connection between the Chern number and the fluctuation-induced Abraham angular momentum. In particular, both the Chern number and the angular momentum spectral density can be expressed in terms of an integral of the photonic Green function along a semi-straight line parallel to the imaginary frequency axis [27]. By exploiting this connection, I show that in a band-gap the angular momentum "quantum" *is* the Chern number. This result together with the theory of Ref. [23], establish the formal link between the Chern number and the net-number of unidirectional edge states, and thereby demonstrate the bulk-edge correspondence in photonics. I numerically illustrate the application of the developed concepts to a gyrotropic photonic crystal cavity.

In short, the key idea that drives the analysis of the article is that the Chern number is determined by the Green function of a system terminated with periodic boundaries, while the fluctuation-induced angular momentum density is determined by the Green function of a cavity terminated with opaque-type walls, i.e., walls impenetrable by the electromagnetic radiation. In a photonic band-gap, the two Green functions are essentially identical in the bulk region. I prove that due to this property the Chern number and the fluctuation-induced angular momentum are profoundly related.

The article is organized as follows: In Sect. II, I present a quick overview of the topological classification of photonic platforms using the system Green function [27] and of the Hamiltonian-type description of a *dispersive* photonic system (e.g., a topological cavity). Both the fluctuation-induced angular momentum and the Chern number can be written either in terms of the normal modes of the equivalent Hamiltonian, or alternatively, in terms of the system Green function. The two different formalisms are used in my analysis. In Sect. III it is demonstrated that the Green function boundary conditions (periodic vs. "opaque") play a critical role in the Chern number calculation. In Sect. IV, the angular momentum expectation is related to the system Green function, and in particular it is shown that its spectral density is given by an integral of the Green function over a semi-straight line parallel to the imaginary frequency axis. The proof that the Chern number is the angular momentum "quantum" is given in Sect. V. The bulk-edge correspondence is formally demonstrated in Sect. VI and is applied to a photonic crystal cavity. A short summary of the main findings is given in Sect. VII.

## II. Topological classification

The topological classification of Chern-type photonic materials is typically based on the normal modes (eigenstates) of the problem. The Maxwell's equations, on their own, do not provide for a Hermitian-type description of dispersive systems [1, 2, 27, 29, 30]. However, the electrodynamics of *lossless* systems can be modeled by a generalized (augmented) Hermitian problem of the form [27, 29, 30]

$$\hat{L} \cdot \underbrace{\begin{pmatrix} \mathbf{f} \\ \mathbf{Q}^{(1)} \\ \mathbf{Q}^{(2)} \\ \dots \end{pmatrix}}_{\mathbf{Q}} = i \frac{\partial}{\partial t} \underbrace{\begin{pmatrix} \mathbf{M}_\infty & 0 & 0 & \dots \\ 0 & 1 & 0 & \dots \\ 0 & 0 & 1 & \dots \\ \dots & \dots & \dots & \dots \end{pmatrix}}_{\mathbf{M}_g} \cdot \begin{pmatrix} \mathbf{f} \\ \mathbf{Q}^{(1)} \\ \mathbf{Q}^{(2)} \\ \dots \end{pmatrix} + i \underbrace{\begin{pmatrix} \mathbf{j} \\ 0 \\ 0 \\ \dots \end{pmatrix}}_{\mathbf{j}_g}. \tag{1}$$

The state vector $\mathbf{Q} = \begin{pmatrix} \mathbf{f} & \mathbf{Q}^{(1)} & \dots & \mathbf{Q}^{(\alpha)} & \dots \end{pmatrix}^T$ depends on the electromagnetic fields $\mathbf{f} = \begin{pmatrix} \mathbf{E} & \mathbf{H} \end{pmatrix}^T$ and on additional variables ($\mathbf{Q}^{(\alpha)}$) which represent the internal degrees of freedom of the material response [1, 2, 27, 29-33]. Both $\mathbf{f}$ and $\mathbf{Q}^{(\alpha)}$ are six-component vectors. Furthermore, $\mathbf{j} = \begin{pmatrix} \mathbf{j}_e & \mathbf{j}_m \end{pmatrix}^T$ is a six-vector with the electric and magnetic current densities. The operator $\hat{L}$ is an integro-differential operator of the form

$$\hat{L}(\mathbf{r}, -i\nabla) = \hat{L}_0(\mathbf{r}) + \begin{pmatrix} \hat{N} & 0 & \dots \\ 0 & 0 & \dots \\ \dots & \dots & \dots \end{pmatrix}, \quad \text{with } \hat{N} = \begin{pmatrix} 0 & i\nabla \times \mathbf{1}_{3\times 3} \\ -i\nabla \times \mathbf{1}_{3\times 3} & 0 \end{pmatrix}. \tag{2}$$

The precise definition of the multiplication operator $\hat{L}_0(\mathbf{r})$ can be found in Refs. [27, 29, 30]. It depends on the poles and residues of the 6×6 material matrix $\mathbf{M}$ that links the frequency domain fields as follows:

$$\begin{pmatrix} \mathbf{D} \\ \mathbf{B} \end{pmatrix} = \underbrace{\begin{pmatrix} \varepsilon_0 \overline{\varepsilon}(\mathbf{r},\omega) & \frac{1}{c}\overline{\xi}(\mathbf{r},\omega) \\ \frac{1}{c}\overline{\zeta}(\mathbf{r},\omega) & \mu_0 \overline{\mu}(\mathbf{r},\omega) \end{pmatrix}}_{\mathbf{M}(\mathbf{r},\omega)} \cdot \begin{pmatrix} \mathbf{E} \\ \mathbf{H} \end{pmatrix}. \tag{3}$$

For completeness, I admit that the material response may be bianisotropic, i.e., the magneto-electric response terms $\overline{\xi}, \overline{\zeta}$ may be nontrivial. Moreover, in Eq. (1) the matrix $\mathbf{M}_\infty$ stands for the asymptotic high-frequency response of the material, $\mathbf{M}_\infty = \lim_{\omega \to \infty} \mathbf{M}(\mathbf{r},\omega)$. The operator $\hat{H}_g(\mathbf{r},-i\nabla) = \mathbf{M}_g^{-1}(\mathbf{r}) \cdot \hat{L}(\mathbf{r},-i\nabla)$ is Hermitian with respect to the weighted inner product [27, 29, 30]

$$\langle \mathbf{Q}_B | \mathbf{Q}_A \rangle \equiv \int_V \frac{1}{2} \mathbf{Q}_B^* \cdot \mathbf{M}_g(\mathbf{r}) \cdot \mathbf{Q}_A d^3\mathbf{r}. \tag{4}$$

The integration is over the volume of the relevant "cavity".

The topological classification of a *periodic* system is typically done by introducing a Berry curvature. The Berry curvature depends explicitly on the Bloch eigenmodes, $\mathbf{Q}_{n\mathbf{k}}(\mathbf{r}) = \tilde{\mathbf{Q}}_{n\mathbf{k}}(\mathbf{r}) e^{i\mathbf{k}\cdot\mathbf{r}}$, of the augmented problem, with $\mathbf{k} = k_x \hat{\mathbf{x}} + k_y \hat{\mathbf{y}}$ the wave vector and the periodic envelope $\tilde{\mathbf{Q}}_{n\mathbf{k}}$ satisfying

$$\hat{H}_g(\mathbf{r},-i\nabla+\mathbf{k}) \cdot \tilde{\mathbf{Q}}_{n\mathbf{k}} = \omega_{n\mathbf{k}} \tilde{\mathbf{Q}}_{n\mathbf{k}}, \tag{5}$$

where $\omega_{n\mathbf{k}}$ represents the eigenfrequency.

The system may be fully three-dimensional: it must be periodic along the *x* and *y* directions, and closed along the *z*-direction; for example, it may be a *lossless* waveguide-type structure terminated with opaque-type boundaries at the planes $z = d$ and $z = 0$ (the top and bottom waveguide walls), *d* being the height of the waveguide (see Fig. 1a). Assuming the normalization $\langle \mathbf{Q}_{n\mathbf{k}} | \mathbf{Q}_{n\mathbf{k}} \rangle = 1$, the Berry curvature of the *n*-th band is [34]

$$\mathcal{F}_{n\mathbf{k}} = i\left[\langle \partial_1\tilde{\mathbf{Q}}_{n\mathbf{k}} | \partial_2\tilde{\mathbf{Q}}_{n\mathbf{k}}\rangle - \langle \partial_2\tilde{\mathbf{Q}}_{n\mathbf{k}} | \partial_1\tilde{\mathbf{Q}}_{n\mathbf{k}}\rangle\right]. \tag{6}$$

In the above, $\partial_i = \partial/\partial k_i$ ($i$=1,2) with $k_1 = k_x$ and $k_2 = k_y$. The gap Chern number is the integral of the Berry curvature over the first Brioullin zone (B.Z.):

$$\mathcal{C} = \frac{1}{2\pi}\iint_{B.Z.} d^2\mathbf{k} \sum_{n\in F} \mathcal{F}_{n\mathbf{k}}. \tag{7}$$

The summation is over all the "filled" photonic bands ($F$) below the gap ($\omega_{n\mathbf{k}} < \omega_{gap}$), with $\omega_{gap}$ some frequency in the relevant band gap. In a band-gap the propagation along directions parallel to the *xoy* plane is forbidden. Note that the modes and band structure depend on the boundary conditions on the top and bottom walls, and hence the band-gaps also do. In particular, the top and bottom walls *cannot* support edge-states in a band-gap. In photonic systems the summation must include both positive and negative frequency branches, because the sum of the Chern numbers of negative frequency bands may be nonzero [27, 35, 36].

Importantly, the topological classification of a photonic system can be done without any detailed knowledge of the photonic band structure or of the Bloch waves, and without introducing a gauge dependent-Berry potential [27, 37]. Indeed, the gap Chern number can be simply expressed as an integral in the complex frequency plane of the photonic Green function $\overline{\mathbf{G}}(\mathbf{r},\mathbf{r}',\omega)$, as follows [27, 38]:

$$\mathcal{C} = \frac{1}{A_{tot}}\operatorname{Re}\int_{\omega_{gap}}^{\omega_{gap}+i\infty} d\omega\, f(\omega), \tag{8}$$

$$f(\omega) = \iint dVdV'\left[\operatorname{tr}\{\partial_2\hat{N}\cdot\overline{\mathbf{G}}(\mathbf{r},\mathbf{r}',\omega)\cdot\partial_1\hat{N}\cdot\partial_\omega\overline{\mathbf{G}}(\mathbf{r}',\mathbf{r},\omega)\} - 1\leftrightarrow 2\right], \tag{9}$$

with $\partial_\omega \equiv \partial/\partial\omega$, tr{...} is the trace operator and $\partial_i \hat{N}$ stands for the 6×6 matrix $\begin{pmatrix} 0 & -\hat{\mathbf{u}}_i \times \mathbf{1}_{3\times 3} \\ \hat{\mathbf{u}}_i \times \mathbf{1}_{3\times 3} & 0 \end{pmatrix}$ with $\hat{\mathbf{u}}_i$ a unit vector along the *i*-th direction. The term "1 ↔ 2" stands for the first term in rectangular brackets with the indices "1" and "2" interchanged. This notation will be used throughout the article. It is worth pointing out that in the electronic case the Chern number is also determined by the fermionic Green function [39]. The integration in Eq. (8) is along the section of the straight line $\text{Re}\{\omega\} = \omega_{\text{gap}}$ (parallel to the complex frequency imaginary axis) in the upper-half plane ("Re" is the real-part operator). The Chern number can also be written as an integral along a line $\text{Re}\{\omega\} = \omega_{\text{gap}}$ [27], but in the present study it is more useful to restrict the integration path to the upper-half plane. The volume integrals are over the entire cavity, which encompasses many elementary cells of the associated photonic crystal. Moreover, it is implicit that the cavity walls perpendicular to the *x* and *y* directions (i.e., the lateral walls) are terminated with periodic boundary conditions. The transverse cross-section of the cavity has area $A_{tot} = L_1 \times L_2$ and the identity (8) holds in the limit $A_{tot} \to \infty$, so that the structure becomes unbounded in the *x* and *y* directions. By definition, the frequency domain photonic Green function $\overline{\mathbf{G}}(\mathbf{r}, \mathbf{r}', \omega)$ satisfies:

$$\hat{N} \cdot \overline{\mathbf{G}}(\mathbf{r}, \mathbf{r}', \omega) = \omega \mathbf{M}(\mathbf{r}, \omega) \cdot \overline{\mathbf{G}}(\mathbf{r}, \mathbf{r}', \omega) + i\mathbf{1}\delta(\mathbf{r} - \mathbf{r}'), \quad (10)$$

with $\mathbf{1} \equiv \mathbf{1}_{6\times 6}$ and periodic boundary conditions over the lateral walls.

## III. Boundaries matter

The Chern number written as in Eq. (7) depends on the Berry curvature, and thereby on the Bloch eigenmodes envelopes. Thus, it is essential that the considered system is periodic in the *x* and *y*-directions. In contrast, Eq. (8) gives the Chern number in terms of the photonic Green function. As previously mentioned, it is implicit (and necessary so that Eqs. (7) and (8) give the same result) that the system Green function satisfies periodic boundary conditions over the cavity lateral-walls [27].

Is the value of $\mathcal{C}$ given by Eq. (8) in the limit $A_{tot} \to \infty$ sensitive to the boundary conditions enforced on the lateral walls? For example, suppose that the lateral walls are taken as perfect electric conductors (PEC). Will the value of $\mathcal{C}$ change in the limit of an infinitely large ($A_{tot} \to \infty$) cavity, compared to the case wherein the boundary conditions are periodic?

To address these questions, first I note that the bulk photonic crystal does not support any states for a frequency in the integration path ($\text{Re}\{\omega\} = \omega_{gap}$, $\text{Im}\{\omega\} > 0$). Indeed, both in the band-gap and for complex-valued frequencies there are no propagating waves and thereby standing waves cannot be formed in the cavity; it is underlined that the top and bottom walls cannot support any type of edge-states in a band-gap because the band-structure takes into account the effect of these boundaries. This indicates that the Green function value must be nearly unaffected by the lateral-walls boundary conditions when the observation and source points $(\mathbf{r}, \mathbf{r}')$ are *interior* to the cavity. Thus, at first sight it seems that the integral in the right-hand side of (8) should be insensitive to the lateral boundary conditions.

Surprisingly, it is shown in Appendix A that this heuristic understanding is wrong. In particular, for a nontrivial topology one has:

$$\mathcal{C}_{\text{Berry}} = \mathcal{C}_{\text{p}} \neq 0 = \mathcal{C}_{\text{PEC}}, \qquad (11)$$

where $\mathcal{C}_{\text{Berry}}$ is calculated with Eq. (7), $\mathcal{C}_{\text{p}}$ with Eq. (8) with periodic boundary conditions and $\mathcal{C}_{\text{PEC}}$ also with Eq. (8) but with PEC boundary conditions. I will say that the lateral walls are "opaque" when it is possible to guarantee that the value of $\mathcal{C}$ calculated with Eq. (8) vanishes. It is shown in Appendix A that if the boundary conditions ensure that the parameters (in the following $x_1 \equiv x$ and $x_2 \equiv y$ and $\mathbf{Q}_n, \mathbf{Q}_m$ are generic state vectors)

$$R^{(j)} = \frac{1}{i}\left[\left\langle \hat{H}_g \mathbf{Q}_n \mid x_j \mid \mathbf{Q}_m \right\rangle - \left\langle \mathbf{Q}_n \mid \hat{H}_g x_j \mid \mathbf{Q}_m \right\rangle\right], \qquad (j=1,2) \qquad (12)$$

vanish, i.e., that

$$R^{(j)} = 0, \qquad \text{(for opaque-type lateral walls)} \qquad (13)$$

then the lateral-walls are "opaque". Note that because $\hat{H}_g$ is a Hermitian operator one might think that $R^{(j)} = 0$ is a universally valid property. However, as discussed in Appendix A, such an understanding is wrong and for periodic-type boundaries $R^{(j)} \neq 0$. Besides the PEC walls already mentioned, another simple example of an opaque-type boundary is the case of perfect magnetic conductor (PMC) walls. As further detailed in Appendix A, the physical reason for the dissimilar results in Eq. (11) is that for opaque-type boundaries the flux of the Poynting vector must vanish at each individual lateral wall. In contrast, periodic boundaries are *cyclic* (reentrant) and hence the Poynting vector flux may be non-trivial at each individual wall.

When the Green function satisfies opaque-type boundary conditions the integral in Eq. (8) vanishes, notwithstanding the Green function in the interior of the cavity is coincident, for all purposes, with the Green function calculated with periodic boundary conditions for which the integral evaluates to $\mathcal{C}_\mathrm{p} = \mathcal{C}_\mathrm{Berry}$, which generally is nonzero.

Something dramatic must happen at the opaque-type boundaries to justify the dissimilar result (11). The simplest way of explaining why the contribution of the boundary region can be comparable to that of the bulk-volumetric region in Eq. (8) is that the integrand has some singularity when either the observation or the source points approach the boundary. On physical grounds, the resonant response can only occur for real-valued frequencies. Thus, the most sensible way to justify (11) is that for opaque-type boundaries the system supports natural modes with $\omega_n \approx \omega_\mathrm{gap}$. Since bulk states are not allowed (as previously noted, edge-states on the top and bottom walls are also not allowed), the relevant natural modes are necessarily edge-states propagating attached to the lateral-walls. In other words, opaque-type boundaries must close the band-gap of the bulk region. Thus, the previous analysis indicates that a topologically nontrivial photonic system ($\mathcal{C}_\mathrm{Berry} \neq 0$) terminated with opaque-type boundary conditions must support edge-states on the lateral walls. This is the first encounter with the "bulk-edge correspondence" principle. Later, the outlined arguments will be made rigorous.

## IV. Angular momentum

Let us now consider a nontrivial Chern-type insulator *cavity* with opaque-type lateral walls. For now, it is assumed that the system is perfectly isolated from the external environment so that there is no dissipation, i.e., both the cavity walls and the topological material are lossless. Furthermore, it supposed that the electromagnetic fields have no quanta so the system is in the ground state ("quantum vacuum"). A discussion of the thermal states in a weakly dissipative cavity will be presented in Sect. V.B.

The unidirectional nature of the edge states in topological materials implies that the fluctuation-induced light may be characterized by a nontrivial angular momentum. This property was first discussed in Ref. [24], where it was argued that topological systems in *equilibrium* with a thermal bath (or in the quantum vacuum state) may enable the circulation of a heat current in closed orbits. Furthermore, such an effect can in principle be observed in other nonreciprocal (but not necessarily topological) systems [23, 25, 26, 41].

Importantly, it was demonstrated in Ref. [23] that in the continuum limit ($A_{tot} \to \infty$) the spectral-density of the Abraham light-angular momentum ($\mathcal{L}_{T,\omega}$) in a closed cavity in equilibrium with a thermal reservoir at temperature $T$ satisfies

$$\frac{\mathcal{L}_{T,\omega}}{A_{tot}} = \mathcal{E}_{T,\omega} \frac{1}{c^2 \pi} \sum_{\omega_m = \omega} s_m, \qquad \text{with } \mathcal{E}_{T,\omega} = \frac{\hbar \omega}{2} \coth\left(\frac{\hbar \omega}{2 k_B T}\right) \qquad (14)$$

the mean energy of a quantum-harmonic oscillator at temperature $T$ [40]. This is the result discussed in the Introduction and heuristically justified using an analogy with a circular transmission line. In particular, the formula holds with $T = 0$ in the "quantum vacuum case" when the material and the cavity walls are lossless. The finite temperature result assumes vanishingly small (but nonzero) material absorption, e.g., in the cavity walls (see Sect. V.B). It is implicit that at the frequency of interest the system does not support *bulk* states (electromagnetic band-gap). The sum in Eq. (14) is over all the edge-modes and simply counts the difference between the number of edge modes associated with an anti-clockwise power flow ($s_m = +1$) and the number of edge modes associated with a clockwise power flow ($s_m = -1$). The enunciated result is universal: it holds independently if the system is topological or not. In particular, the angular momentum spectral density per unit of area due to thermal or quantum fluctuations is quantized in units of $\mathcal{E}_{T,\omega} \frac{1}{c^2 \pi}$ [23].

Equation (14) links the angular momentum spectral density with the number of edge states. This suggests that for topological systems $\mathcal{L}_{T,\omega}$ may be directly written in terms of the topological Chern number [23]. Next, I develop the theoretical formalism necessary to prove that that is indeed the case.

### A. Classical states

There are two relevant light momenta: the Abraham momentum (kinetic momentum of light) and the Minkowski momentum (canonical momentum of light) [41-44]. This article uses the Abraham formalism, which leads to a quantization of the angular momentum spectral density in the topological cavity. The fluctuation-induced Minkowski angular spectral density is not quantized. For discussion on the detailed meaning of each momenta the reader is referred to Refs. [41, 44].

The Abraham (kinetic) light momentum in the cavity is determined by $\frac{1}{c^2} \int dV \, \mathbf{S}$ where $\mathbf{S} = \mathrm{Re}\{\mathbf{E} \times \mathbf{H}^*\}$ the Poynting vector of a complex-valued field [41, 42, 43]. From Appendix A, the *i*-th component of the momentum can be written as [Eq. (A5)]:

$$\frac{1}{c^2} \hat{\mathbf{u}}_i \cdot \int dV \, \mathbf{S} = \left\langle \mathbf{Q} \left| \frac{1}{c^2} \partial_i \hat{H}_g \right| \mathbf{Q} \right\rangle, \qquad (15)$$

where $\partial_i \hat{H}_g = \dfrac{\partial}{\partial k_i}\left[\hat{H}_g(\mathbf{r}, -i\nabla + \mathbf{k})\right]$. Thus, $\dfrac{1}{c^2}\partial_\mathbf{k}\hat{H}_g$ may be understood as the Abraham momentum operator. From Appendix A, in the photonic case $\partial_\mathbf{k}\hat{H}_g$ is independent of the wave vector. Then, it follows that the (kinetic) angular momentum operator is $\hat{\mathcal{L}} = \dfrac{1}{c^2}\mathbf{r} \times \partial_\mathbf{k}\hat{H}_g$ (the nomenclature of Ref. [45] is adopted here). In particular, the $z$-component of the angular momentum (perpendicular to the plane of propagation) of a given state vector $\mathbf{Q}$ is:

$$\mathcal{L}_z = \langle \mathbf{Q} | \hat{\mathcal{L}}_z | \mathbf{Q} \rangle = \dfrac{1}{c^2}\langle \mathbf{Q} | x_1 \partial_2 \hat{H}_g - x_2 \partial_1 \hat{H}_g | \mathbf{Q} \rangle. \tag{16}$$

The angular momentum can be expressed explicitly in terms of the electromagnetic fields as $\mathcal{L} = \dfrac{1}{c^2}\int dV\, \mathbf{r} \times \mathbf{S}$ [23]. The angular momentum of light is extensively discussed in Refs. [45-48]. As further detailed in Appendix A, $\partial_\mathbf{k}\hat{H}_g$ is given by the commutator of the position operator and the pseudo-Hamiltonian: $\partial_\mathbf{k}\hat{H}_g = \dfrac{1}{i}\left[\mathbf{r}, \hat{H}_g\right]$ [Eq. (A4)]. Thus, the angular momentum may also be written as:

$$\mathcal{L}_z = \dfrac{1}{c^2} i \langle \mathbf{Q} | x_1 \hat{H}_g x_2 - x_2 \hat{H}_g x_1 | \mathbf{Q} \rangle. \tag{17}$$

### *B. Quantum vacuum state*

In a lossless dispersive material cavity the electromagnetic field can be quantized by letting each normal classical mode become a quantum harmonic oscillator [49]. The angular momentum of the "quantum vacuum" ($\langle \mathcal{L}_z \rangle_{T=0^+}$) can be easily found noting that the energy stored in the $n$-th mode is determined by the zero-point energy

-16-

$\mathcal{E}_{0,\omega_n} = \hbar|\omega_n|/2$. Hence, adding-up the contributions of all (positive-frequency) modes one finds that the expectation of the light angular momentum in the cavity is:

$$\langle \mathcal{L}_z \rangle_{T=0^+} = \sum_{\omega_n>0} \mathcal{E}_{0,\omega_n} \mathcal{L}^{(n)}, \qquad \text{with } \mathcal{L}^{(n)} \equiv \frac{i}{c^2} \frac{\langle \mathbf{Q}_n | x_1 \hat{H}_g x_2 - x_2 \hat{H}_g x_1 | \mathbf{Q}_n \rangle}{\langle \mathbf{Q}_n | \mathbf{Q}_n \rangle}. \qquad (18)$$

It is supposed that the modes are normalized such that $\langle \mathbf{Q}_n | \mathbf{Q}_m \rangle = \delta_{n,m}$. Note that $\langle \mathbf{Q} | \mathbf{Q} \rangle$ gives the stored energy [27, 30] and thus the parameter $\mathcal{L}^{(n)}$ has unities of angular momentum per Joule. Similar to Ref. [24], it is simple to show that Eq. (18) can be directly obtained from the fluctuation-dissipation theorem using a modal expansion of the system Green function.

In open systems, the total light-angular momentum may depend on the origin of the coordinate axes [45]. Importantly, for a closed cavity the expectation of the *total* light momentum vanishes (even though locally the light momentum density is generally nontrivial [24]). Indeed, for opaque-type walls ($R^{(i)} = 0$), Eq. (A6) shows that a generic cavity mode satisfies:

$$\langle \mathbf{Q}_n | \partial_i \hat{H}_g | \mathbf{Q}_n \rangle = 0, \qquad (i=1,2). \qquad (19)$$

Combining this result with Eq. (16), it follows that $\mathcal{L}^{(n)}$ is origin independent. Thereby, the expectation of $\mathcal{L}_z$, i.e., of the total angular momentum, is also origin independent.

I introduce a (unilateral) "quantum vacuum" angular momentum spectral density $\mathcal{L}_\omega$ such that $\langle \mathcal{L}_z \rangle_{T=0^+} = \int_0^\infty d\omega\, \mathcal{L}_\omega$. Clearly, it has the modal expansion [23]:

$$\mathcal{L}_\omega = \mathcal{E}_{0,\omega} \sum_{\omega_n>0} \mathcal{L}^{(n)} \delta(\omega - \omega_n). \qquad (20)$$

Using the fluctuation-dissipation theorem [40] with $T = 0^+$, (which is applicable to the ground state of lossless closed systems, see [24]) it is proven in Appendix B that $\mathcal{L}_\omega$ can alternatively be written in terms of the system Green function of the lossless cavity:

$$\mathcal{L}_\omega = -\mathcal{E}_{0,\omega} \frac{1}{c^2} \frac{1}{\pi} \operatorname{Re}\left\{ \int dV \operatorname{tr}\left\{ \overline{\mathbf{G}}(\mathbf{r},\mathbf{r},\omega) \cdot \left( x_1 \partial_2 \hat{N} - x_2 \partial_1 \hat{N} \right) \right\} \bigg|_{\omega + 0^+ i} \right\}. \tag{21}$$

### C. "Partial" angular momentum

It is useful to introduce a partial "quantum-vacuum" angular-momentum, $\mathcal{L}_z(\omega_g)$, including only the contributions of the modes with $\omega_n > \omega_g > 0$. Since the zero-point energy of a harmonic oscillator is $\mathcal{E}_{0,\omega} = \hbar|\omega|/2$, the partial angular momentum is given by (compare with Eq. (18)):

$$\mathcal{L}_z(\omega_g) \equiv \sum_{\omega_n > \omega_g} \frac{\hbar \omega_n}{2} \mathcal{L}^{(n)}, \tag{22}$$

Clearly, the quantum-vacuum angular momentum satisfies $\langle \mathcal{L}_z \rangle_{T=0^+} = -\int_{0^+}^{\infty} d\omega_g \frac{d\mathcal{L}_z}{d\omega_g} = \mathcal{L}_z(\omega_g = 0^+)$. Thus, the spectral density is:

$$\mathcal{L}_\omega = -\frac{d\mathcal{L}_z}{d\omega_g}(\omega). \tag{23}$$

By integrating the fluctuation-dissipation theorem result (21) in the interval $\omega_g < \omega < +\infty$, one obtains, after performing a Wick rotation centered at $\omega = \omega_g$, the following alternative formula for $\mathcal{L}_z(\omega_g)$:

$$\mathcal{L}_z(\omega_g) = -\frac{1}{c^2} \frac{\hbar}{2\pi} \operatorname{Re}\left\{ \int dV \int_{\omega_g + 0^+ i}^{\omega_g + i\infty} d\omega\, \omega\, \operatorname{tr}\left\{ \overline{\mathbf{G}}(\mathbf{r},\mathbf{r},\omega) \cdot \left( x_1 \partial_2 \hat{N} - x_2 \partial_1 \hat{N} \right) \right\} \right\}. \tag{24}$$

The integration path is along the section of the straight line $\text{Re}\{\omega\} = \omega_g$ in the upper-half plane.

Interestingly, the partial-angular momentum $\mathcal{L}_z(\omega_g)$ can be expressed in terms of the function $f$ defined by Eq. (9). Specifically, for a cavity with opaque-type lateral walls one has:

$$\mathcal{L}_z(\omega_g) = \frac{1}{c^2}\frac{-\hbar}{4\pi}\text{Re}\int_{\omega_g}^{\omega_g+i\infty} d\omega\, \omega^2 f(\omega). \qquad (25)$$

I present two independent derivations of this formula: in Appendix C using the fluctuation-dissipation theorem result [Eq. (24)] and in Appendix D using the modal expansion (22). Since $\mathcal{L}_z(\omega_g = 0^+)$ represents the "quantum-vacuum" angular momentum expectation, it follows that $\langle \mathcal{L}_z \rangle_{T=0^+}$ can be written as an integral of the system Green function over the imaginary frequency axis. This is analogous to Casimir's theory where the zero-point energy of a system is determined by an integral over "imaginary" frequencies [50, 51]. The angular momentum expectation is manifestly independent of the coordinate system origin.

### D. Cavity with periodic lateral walls

Up to now, in this section it was assumed that the lateral walls of the cavity are "opaque". However, Eqs. (23) and (25) can be readily extended (from a mathematical standpoint) to a cavity terminated with periodic boundaries. Does the result calculated with Eqs. (23) and (25) in the limit of an infinitely large ($A_{tot} \to \infty$) cavity depend on the Green function boundary conditions when $\omega$ lies in a band-gap of the bulk region? Similar to Sect. III, it turns out that even though $\mathcal{L}_\omega$ is determined by a volume integral it critically depends on the boundary conditions. Specifically, it is shown in Appendix D that for periodic lateral boundaries:

$$\left.\mathcal{L}_\omega\right|_{\omega=\omega_{gap}} = 0, \quad \text{(periodic boundaries)} \quad (26)$$

when $\omega_{gap}$ is in a band-gap of the bulk-region. In contrast, for opaque-type walls typically one has $\left.\mathcal{L}_\omega\right|_{\omega=\omega_{gap}} \neq 0$ [Eq. (14)]. Equation (26) is consistent with the fact that the bulk material does not support any states in the band-gap and hence its angular momentum density must vanish.

## V. Quantized angular momentum spectral density

### A. Quantum fluctuations

Let us introduce $\mathcal{C}_\omega^\mathcal{L}$ defined such that the expectation of the "quantum-vacuum" angular momentum spectral density per unit of area satisfies:

$$\frac{\mathcal{L}_\omega}{A_{tot}} = -\frac{1}{\pi c^2} \mathcal{E}_{0,\omega} \mathcal{C}_\omega^\mathcal{L}. \quad (27)$$

The function $\mathcal{C}_\omega^\mathcal{L}$ is dimensionless and from Eqs. (23) and (25), it may be generally written as:

$$\mathcal{C}_\omega^\mathcal{L} = \frac{-1}{2\omega} \frac{d}{d\omega_g} \left[ \frac{1}{A_{tot}} \text{Re} \int_{\omega_g}^{\omega_g + i\infty} d\omega\, \omega^2 f(\omega) \right]. \quad (28)$$

Clearly, the value of $\mathcal{C}_\omega^\mathcal{L}$ depends critically on the Green function boundary conditions. Let $\overline{\mathbf{G}}_p$ be the Green-function calculated with periodic boundaries and let $\overline{\mathbf{G}}_o$ be the Green-function calculated with the relevant opaque-type boundaries (e.g., PEC boundaries). As discussed in Sect. III, for observation and source points *interior* to the cavity one has

$$\overline{\mathbf{G}}_{\mathrm{p}}(\mathbf{r},\mathbf{r}',\omega) \approx \overline{\mathbf{G}}_{\mathrm{o}}(\mathbf{r},\mathbf{r}',\omega), \tag{29}$$

when $\mathrm{Re}\{\omega\}$ is in a band-gap of the bulk region. Furthermore, the larger is $\mathrm{Im}\{\omega\} > 0$ the better is the approximation because the Green function becomes more localized in space for a large $\mathrm{Im}\{\omega\}$. Let us denote $f_l(\omega)$ the function defined by Eq. (9) with $\overline{\mathbf{G}} = \overline{\mathbf{G}}_l$, $l$=p,o. Then, in the limit $A_{tot} \to \infty$ it is possible to write:

$$\begin{aligned} \mathcal{C}_\omega^{\mathcal{L}}\bigg|_{\omega=\omega_{\mathrm{gap}}} &= \frac{-1}{2\omega_{\mathrm{gap}}} \frac{d}{d\omega_{\mathrm{gap}}} \left[ \frac{1}{A_{tot}} \mathrm{Re} \int_{\omega_{\mathrm{gap}}}^{\omega_{\mathrm{gap}}+i\infty} d\omega \, f_{\mathrm{o}}(\omega) \omega^2 \right] \\ &= \frac{-1}{2\omega_{\mathrm{gap}}} \frac{d}{d\omega_{\mathrm{gap}}} \left[ \frac{1}{A_{tot}} \mathrm{Re} \int_{\omega_{\mathrm{gap}}}^{\omega_{\mathrm{gap}}+i\infty} d\omega \left[ f_{\mathrm{o}}(\omega) - f_{\mathrm{p}}(\omega) \right] \omega^2 \right] \\ &\approx \frac{-1}{2\omega_{\mathrm{gap}}} \frac{d}{d\omega_{\mathrm{gap}}} \left[ \omega_{\mathrm{gap}}^2 \frac{1}{A_{tot}} \mathrm{Re} \int_{\omega_{\mathrm{gap}}}^{\omega_{\mathrm{gap}}+i\infty} d\omega \left[ f_{\mathrm{o}}(\omega) - f_{\mathrm{p}}(\omega) \right] \right] \end{aligned} \tag{30}$$

The second identity is a consequence that the term with the periodic Green function vanishes [see Eq. (26)], while the third identity is due to the fact that for $A_{tot} \to \infty$ one can safely assume that $f_{\mathrm{o}}(\omega) \approx f_{\mathrm{p}}(\omega)$ in the integration path, except in the immediate vicinity of the real-frequency axis where $\omega \approx \omega_{\mathrm{gap}}$ [52]. But from the results of Sect. III and Eq. (8), it is known that $\frac{1}{A_{tot}} \mathrm{Re} \int_{\omega_{\mathrm{gap}}}^{\omega_{\mathrm{gap}}+i\infty} d\omega \, f_{\mathrm{o}}(\omega) = 0$ (the Chern number vanishes with opaque-type boundaries), whereas $\frac{1}{A_{tot}} \mathrm{Re} \int_{\omega_{\mathrm{gap}}}^{\omega_{\mathrm{gap}}+i\infty} d\omega \, f_{\mathrm{p}}(\omega) = \mathcal{C}$, with $\mathcal{C}$ the system Chern number (calculated with periodic boundaries and $A_{tot} \to \infty$). These results imply that

-21-

$\left.\mathcal{C}_\omega^\mathcal{L}\right|_{\omega=\omega_{\text{gap}}} = \frac{1}{2\omega_{\text{gap}}} \frac{d}{d\omega_{\text{gap}}}\left[\omega_{\text{gap}}^2 \mathcal{C}\right]$. The Chern number $\mathcal{C}$ is independent of $\omega_{\text{gap}}$. Therefore, one finally obtains

$$\left.\mathcal{C}_\omega^\mathcal{L}\right|_{\omega=\omega_{\text{gap}}} = \mathcal{C}, \qquad \text{(in a gap of the bulk-states).} \qquad (31)$$

Thus, in agreement with Ref. [23], I conclude that the quantum fluctuation-induced angular momentum in a topological system is precisely quantized. Its "quantum" is exactly the gap Chern number. It is underlined that the derivation of Eq. (31) is fully independent of Ref. [23].

## *B. Thermal fluctuations*

The results of Sect. IV.B can be readily generalized to weakly dissipative systems, e.g., a cavity filled with the topological material with the cavity walls slightly absorptive. Such systems are coupled to the external environment (e.g., through the cavity walls) and hence support thermal states. The thermal states of weakly dissipative systems can be studied perturbatively simply by considering that the mean energy of the *n*-th mode is $\mathcal{E}_{T,\omega}$. For example, the angular momentum expectation at temperature $T$ is $\langle \mathcal{L}_z \rangle_T \approx \sum_{\omega_n > 0} \mathcal{E}_{T,\omega_n} \mathcal{L}^{(n)}$, with $\mathcal{L}^{(n)}$ evaluated using the modes of the corresponding idealized lossless cavity (with perfectly reflecting walls). Within this approximation, the spectral density [Eq. (20)] should evidently be replaced by $\mathcal{L}_{T,\omega} \approx \mathcal{E}_{T,\omega} \sum_{\omega_n > 0} \mathcal{L}^{(n)} \delta(\omega - \omega_n)$. Thereby, it follows that in the outlined conditions $\mathcal{L}_{T,\omega}/\mathcal{E}_{T,\omega}$ is independent of the temperature and in particular, $\mathcal{L}_{T,\omega}/\mathcal{E}_{T,\omega} \approx \mathcal{L}_\omega/\mathcal{E}_{0,\omega}$. Thereby, from Eqs. (27) and (31) the thermally-induced angular momentum per unit of area is also quantized in units of $\mathcal{E}_{T,\omega}\frac{1}{c^2\pi}$:

$$\frac{\mathcal{L}_{T,\omega}}{A_{tot}}=-\frac{1}{\pi c^2}\mathcal{E}_{T,\omega}\mathcal{C}. \tag{32}$$

## VI. Proof of the bulk-edge correspondence

Combining Eq. (32) with Eq. (14) (derived in Ref. [23]), one sees that in a topological system the gap Chern number is linked to the net-number of unidirectional edge states as:

$$\mathcal{C}=-\sum_{\omega_m=\omega}s_m. \tag{33}$$

Thus, the Chern number of the bulk region determines precisely the net number of edge modes circulating around the lateral "opaque-type" walls of the closed cavity. In particular, a nontrivial Chern number implies the emergence of unidirectional gapless edge-modes. For a positive (negative) gap Chern number the unidirectional modes propagate clockwise (anti-clockwise) with respect to the z-axis.

This result may be further generalized to give the number of edge modes propagating at the interface of two topological materials: "the bulk-edge correspondence". To that end, I consider the geometry depicted in Fig. 2, which shows a cavity half-filled with two photonic insulators (the two materials share a photonic band-gap). The cavity lateral walls are assumed "opaque". Let $\mathcal{C}_1$ and $\mathcal{C}_2$ be the gap Chern numbers for material 1 and 2, respectively. Formula (33) implies that $\mathcal{C}_1$ and $\mathcal{C}_2$ determine the number of modes propagating in the clockwise direction around the cavity walls (see Fig. 2). Hence, the number of modes propagating at the interface of the two materials (along the $+x_1$ direction) must be precisely $C_2 - C_1$, i.e., the gap Chern number difference.

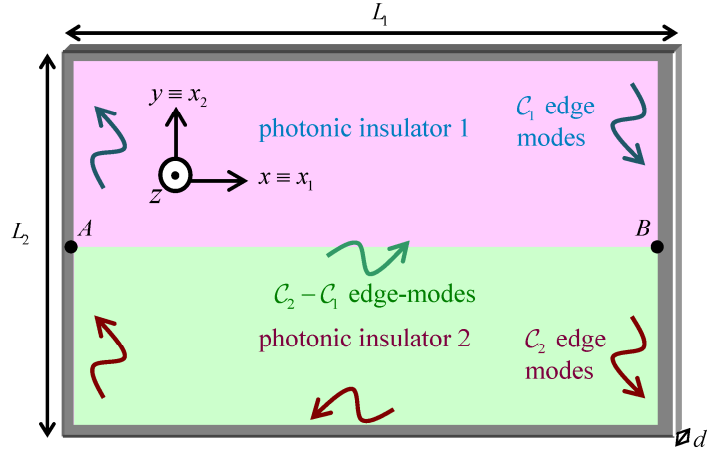

Fig. 2 Illustration of the bulk-edge correspondence principle. A cavity (terminated with "opaque-type" lateral walls) is filled with two photonic insulators. The points $A$ and $B$ represent the two junctions.

The reason why this needs to be so is that otherwise the system would be unstable and a steady-state could not be reached. Indeed, suppose that the net number of unidirectional modes propagating at the interface of the two materials is different from $C_2 - C_1$. In this situation, for one of the junction points (let us say point B of Fig. 2) the number of edge modes arriving at the junction is larger than the number of edge-modes propagating away from the junction. Since the system response is linear, this implies that it would be possible to choose the complex amplitudes of the incident waves in such a way that the edge waves propagating away from the junction are not excited. But then, since by assumption there is no loss and there are not scattering channels available, the energy incident in the junction must remain stored in it. Hence, it is impossible to reach a steady-state for a time-harmonic excitation: the energy stored at the junction grows linearly with time similar to a lossless LC circuit excited at the resonance. Physically this is not acceptable, and hence the net number of unidirectional edge-modes propagating at the interface of the two-materials must be precisely $C_2 - C_1$. This concludes the proof of the bulk-edge correspondence principle.

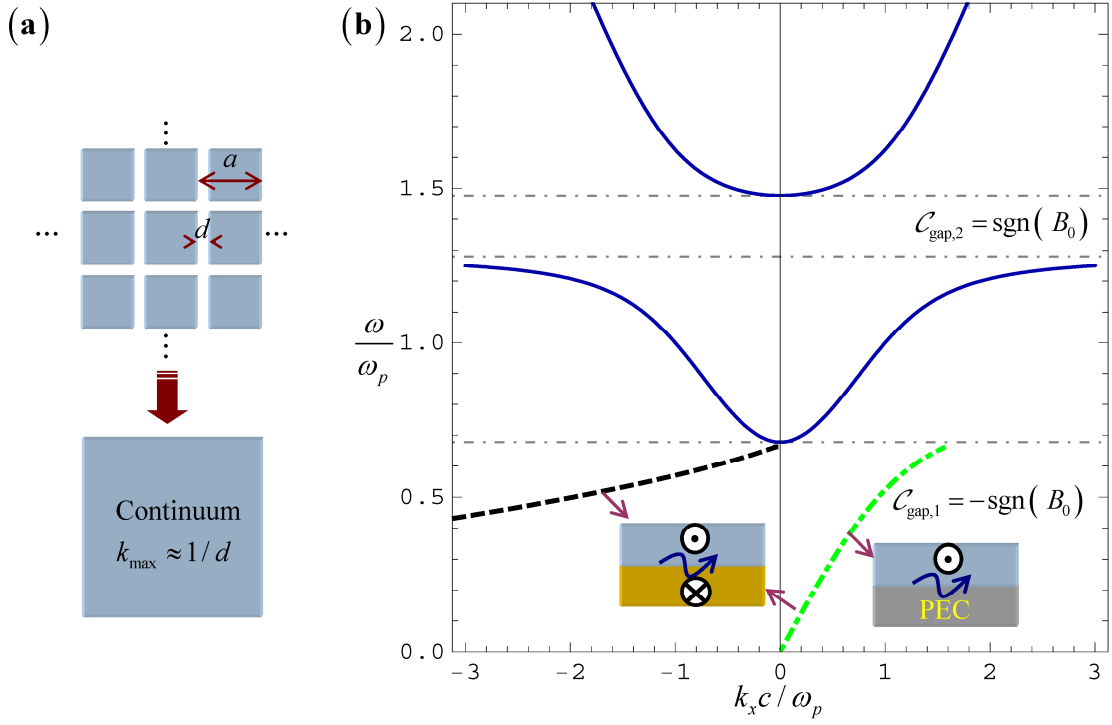

Fig. 3 **(a)** Geometry of a two-dimensional photonic crystal formed by squared-shaped gyrotropic-material inclusions organized in a square lattice. For sufficiently low frequencies the photonic crystal may be regarded as a continuum with a spatial-frequency cut-off $k_{max}$. **(b)** Band structure of material (solid blue lines) and dispersion of the edge-states in the 1$^{st}$ band-gap for *(i)* a gyrotropic-PEC interface (dot-dashed green line) and *(ii)* gyrotropic-gyrotropic interface with the materials biased with magnetic fields oriented in opposite directions (dot-dashed green line and dashed black line). The band-structure, the edge-state dispersions and the gap Chern-numbers (indicated in the insets) are found using the continuum approximation.

To illustrate the application of the developed theory, I consider a two-dimensional photonic crystal (the condition $\partial/\partial z = 0$ in enforced) formed by squared shaped nonreciprocal inclusions organized in a square lattice with period $a$ (Fig. 3a). The

-25-

inclusions stand in air and are spaced by *d*. Furthermore, the analysis is restricted to transverse-magnetic (TM) polarized waves with nontrivial field components $H_z, E_x, E_y$. The inclusions electric response is assumed to be gyrotropic with the same dispersion model as a lossless magnetized plasma [53] (e.g., a magnetized semiconductor [54]) $\overline{\varepsilon} = \varepsilon_t \mathbf{1}_t + \varepsilon_a \hat{\mathbf{z}} \otimes \hat{\mathbf{z}} + i\varepsilon_g \hat{\mathbf{z}} \times \mathbf{1}$, with

$$\varepsilon_t = 1 - \frac{\omega_p^2}{\omega^2 - \omega_c^2}, \qquad \varepsilon_a = 1 - \frac{\omega_p^2}{\omega^2}, \qquad \varepsilon_g = \frac{1}{\omega}\frac{\omega_c \omega_p^2}{\omega_c^2 - \omega^2}, \tag{34}$$

and $\mathbf{1}_t = \hat{\mathbf{x}} \otimes \hat{\mathbf{x}} + \hat{\mathbf{y}} \otimes \hat{\mathbf{y}}$. In the above, $\omega_p$ is the plasma frequency, $\omega_c = -qB_0/m$ is the cyclotron frequency (positive when the magnetic field is oriented along +z), $q = -e$ is the electron charge and *m* is the electron effective mass [53].

The structural parameters of the photonic crystal are $a = \frac{2\pi}{5}\frac{c}{\omega_p}$ and $d = 0.1a$. For $\omega \ll \omega_p$ the air-gaps are deeply subwavelength, and hence in the long-wavelength limit it seems reasonable to approximate the photonic crystal by a continuum with the same permittivity as the inclusions, as illustrated in Fig. 3a. This approximation greatly simplifies the calculation of the band structure and of the gap Chern numbers. In order that the electromagnetic continuum is topological it is necessary to impose a high-frequency spatial cut-off, $k_{\max}$ [29]. For the physical reasons discussed in detail in Ref. [15], the spatial cut-off should be taken on the order of $k_{\max} \approx 1/d$. The photonic band structure obtained with the continuum approximation is depicted in Fig. 3b (solid blue lines) for $\omega_c = \pm 0.8\omega_p$. As seen, there are two band-gaps and the corresponding gap Chern numbers are indicated in the insets. The Chern numbers calculation is done as in Ref. [27] and takes into account the contribution of the negative frequency bands (not shown in Fig. 3b).

Next, I focus on the low-frequency band-gap for which the continuum approximation is arguably more accurate. Its gap Chern number is $\mathcal{C}_{\text{gap},1} = -\text{sgn}(B_0) = -\text{sgn}(\omega_c)$, and thus it is topologically nontrivial. Hence, if the material is paired with a PEC boundary, the bulk-edge correspondence predicts that there is a single edge-state propagating along the +x-direction. To confirm this prediction, I used the continuum approximation to compute the edge-states dispersion. The spatial cut-off $k_{\max}$ is taken into account using the spatially dispersive model described in Ref. [15]. The calculated dispersion (for a material biased with $B_0 > 0$ and $\omega_c = 0.8\omega_p$) is plotted with a green-dotted line in Fig. 3b, and yields the unidirectional gapless edge-mode.

It is also interesting to analyze the case in which two topologically distinct plasmas are paired to form an interface (inset of Fig. 3b). In this scenario, the top region ($y > 0$) is biased with $B_0 > 0$ ($\omega_c = +0.8\omega_p$) and the bottom region ($y < 0$) is biased with $B_0 < 0$ ($\omega_c = -0.8\omega_p$). The gap Chern number difference is now $-1-1=-2$, and hence the bulk-edge correspondence predicts two modes propagating along the +x-direction. This property is confirmed by the numerical results: the edge-state dispersion is now formed by two branches. Due to the symmetry of the structure, one of the branches (with $k_x > 0$) is coincident with the one obtained for the gyrotropic-PEC interface geometry discussed previously. The second branch has $k_x < 0$ but a positive group velocity, i.e., it is a backward wave. Thus, in agreement with the bulk-edge correspondence, both edge-modes propagate along the +x-direction.

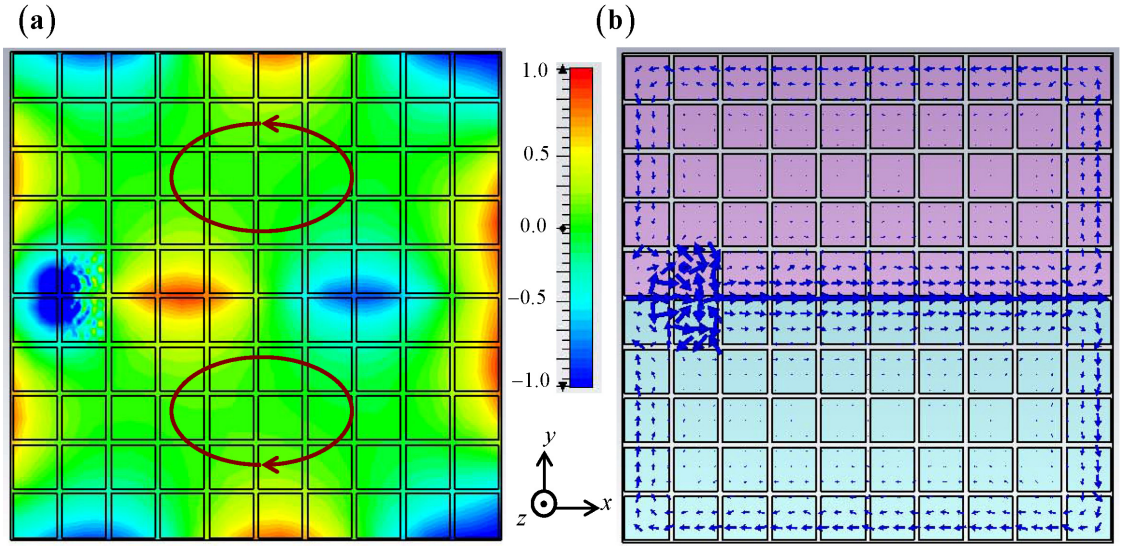

Fig. 4 Photonic crystal cavity terminated with PEC lateral walls. The region $y > 0$ (top-half of the cavity) is biased with a positive (along +z) magnetic field ($\omega_c = +0.8\omega_p$) and the region $y < 0$ (bottom-half of the cavity) with a negative (along –z) magnetic field ($\omega_c = -0.8\omega_p$). The cavity is excited with a vertical (along +y) short-electric dipole placed at the interface of the two regions near the left-hand side lateral wall. The oscillation frequency of the dipole is $\omega = 0.5\omega_p$. **(a)** Time snapshot of the magnetic field $H_z$. **(b)** Poynting vector lines showing how the energy circulates in the cavity.

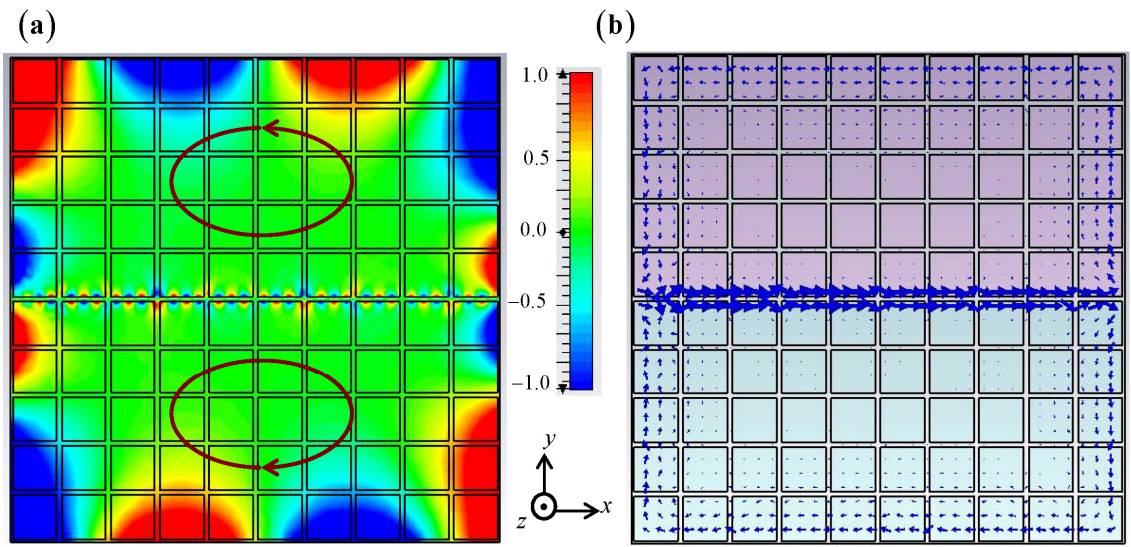

Fig. 5 Similar to Fig. 4 but for a horizontal (along +x) electric dipole.

To further validate the analysis and the link between the angular momentum and the gap Chern numbers, I used CST Microwave Studio [55] to simulate the full wave response of a photonic crystal cavity with a geometry analogous to that of Fig. 2. The cavity lateral walls are PEC. The top region ($y > 0$) is a truncated photonic crystal with $\omega_c = +0.8\omega_p$, and the bottom region ($y < 0$) a truncated photonic crystal with $\omega_c = -0.8\omega_p$. The structural parameters of the photonic crystals are as in the previous example. The CST simulations fully take into account the granular structure of the photonic crystals (the continuum approximation is not used). From the continuum results (Fig. 3), one may expect that for low frequencies this system supports (i) one unidirectional edge-state propagating along the lateral walls and (ii) two distinct unidirectional edge-states propagating along the interface ($y = 0$) of the two gyrotropic photonic crystals. To test these ideas, the cavity was excited with a dipole-type antenna placed in between the two photonic crystals near to the left-hand side lateral wall. Figures 4a and 5a show a time snapshot of the excited magnetic field ($H_z$) for a dipole oriented perpendicular (vertical dipole) and parallel (horizontal dipole) to the interface, respectively, with oscillation frequency $\omega = 0.5\omega_p$. The effect of weak material loss is taken into account to ensure the convergence of the simulations. The propagation of edge-states at the lateral-walls and at the interface of the two-photonic crystals is evident. Furthermore, as can be seen from the Poynting vector lines in Figs. 4b and 5b the energy circulates in closed orbits, such that for the top region (with gap Chern number $\mathcal{C}_{\text{gap},1} = -1$) the energy flows in the anti-clockwise direction whereas in the bottom region (with gap Chern number $\mathcal{C}_{\text{gap},1} = +1$) it flows in the clockwise direction. This result is in agreement with Eq. (33), which links the sign of the Chern number with the direction along which the energy circulates. Interestingly, the time-animations available in the supplementary materials [56] show that the edge-state excited at the interface of the two photonic crystals ($y=0$) by the vertical dipole is a forward wave (Fig. 4), whereas the edge state excited by the horizontal dipole is a backward wave (Fig. 5). Hence, in agreement with the dispersion of the edge-states obtained with the continuum approximation (Fig. 3b), the interface $y=0$ supports two unidirectional edge-modes: a forward wave and a backward wave. Furthermore, as seen in Figs. 4 and 5, $H_z$ has even (odd) symmetry with respect to $y=0$ for the forward (backward) mode, respectively. The edge-mode profiles obtained with the continuum theory have the same symmetries, which further reinforces the validity of this approximation.

## VII. Summary

I established a direct link between the thermal (or quantum) expectation of the light angular momentum spectral density $\mathcal{L}_{T,\omega}$ in a large cavity and the photonic Chern number. Using as a starting point either the fluctuation-dissipation theorem or a modal expansion approach, it was shown that in a cavity with opaque-type boundaries $\mathcal{L}_{T,\omega}$ can be written as an integral of the system Green function over a semi-straight line parallel to the imaginary-frequency axis [Eqs. (23) and (25)]. Taking into account that the gap Chern number $\mathcal{C}$ has a similar representation [Eq. (8)], it was proven that for a sufficiently large cavity ($A_{tot} \to \infty$) and in a band-gap of the bulk region the following universal relation –independent of the material properties and of the photonic crystal

geometry– $\mathcal{L}_{T,\omega} / A_{tot} = -\mathcal{E}_{T,\omega}\mathcal{C}/\pi c^2$ holds. Thus, in agreement with Ref. [23], it follows that the spectral density of the angular momentum per unit of area is quantized in the bulk band-gaps, and that the Chern number is the "quantum" of the fluctuation-induced angular momentum. Furthermore, using the findings of Ref. [23], which link $\mathcal{L}_{T,\omega}$ with the net number of unidirectional edge-modes circulating around the cavity walls, I presented a physically intuitive proof of the bulk-edge correspondence principle in photonics [Eq. (33)].

From a more formal point of view, my theory highlights how topological invariants (with are computed for periodic structures) manifest themselves when the system is *closed* with opaque-type lateral walls. For example, the photonic Chern number manifests itself in the form of a non-trivial fluctuation induced angular-momentum. Moreover, the theory suggests that the recently discovered quadrupole-topological index [9, 57] may be linked to higher-order multipoles of the light-momentum thermal expectation. These higher-order multipoles may represent localized vortices of the heat current. Thus, my findings raise multiple questions about the intriguing role of topology in fluctuation electrodynamics and point to new exciting research directions.

**Acknowledgements:** This work is supported in part by Fundação para a Ciência e a Tecnologia grant number PTDC/EEI-TEL/4543/2014 and by Instituto de Telecomunicações under project UID/EEA/50008/2017.

# Appendix A: Dependence of the Chern number on the boundary conditions

In this Appendix, it is demonstrated that $\mathcal{C}$ calculated with Eq. (8) depends critically on the Green function boundary conditions [Eq. (11)].

## A. Alternative formula for the Chern number

To begin with, I use the fact that *independent* of the boundary conditions enforced on the lateral walls, the value of $\mathcal{C}$ calculated with Eq. (8) is the same as [27, see Ap. B]:

$$\mathcal{C} = \lim_{A_{tot} \to \infty} \frac{2\pi}{A_{tot}} \sum_{n \in F} \mathcal{F}_n, \tag{A1}$$

with $\mathcal{F}_n$ given by:

$$\mathcal{F}_n = \sum_{m \neq n} i \frac{1}{(\omega_n - \omega_m)^2} \left[ \langle \mathbf{Q}_n | \partial_1 \hat{H}_g | \mathbf{Q}_m \rangle \langle \mathbf{Q}_m | \partial_2 \hat{H}_g | \mathbf{Q}_n \rangle - 1 \leftrightarrow 2 \right]. \tag{A2}$$

The sum in Eq. (A1) is restricted to the cavity modes with $\omega_n < \omega_{gap}$, whereas the sum in (A2) is over all the modes. It is relevant to point out that for a bounded cavity ($A_{tot}$ finite) the set of modes is countable.

Furthermore, $\partial_i \hat{H}_g$ represents the operator $\frac{\partial}{\partial k_i}\left[ \hat{H}_g(\mathbf{r}, -i\nabla + \mathbf{k}) \right]$. Unlike in electronics, in photonic systems $\hat{H}_g(\mathbf{r}, -i\nabla + \mathbf{k})$ is a linear function of the wave vector, and hence $\partial_i \hat{H}_g$ is independent of $\mathbf{k}$. From Eq. (2), it can be explicitly written as $\partial_i \hat{H}_g = \mathbf{M}_g^{-1} \cdot \partial_i \hat{L}$ with

$$\partial_i \hat{L} = \begin{pmatrix} \partial_i \hat{N} & \mathbf{0} & \dots \\ \mathbf{0} & \mathbf{0} & \dots \\ \dots & \dots & \dots \end{pmatrix}. \tag{A3}$$

In Eq. (A2) the modes $\{\mathbf{Q}_n\}_{n=1,2,\dots}$ are the full-cavity modes ($\hat{H}_g(\mathbf{r}, -i\nabla) \cdot \mathbf{Q}_n = \omega_n \mathbf{Q}_n$), rather than the periodic spatial envelopes of the Bloch eigenmodes (with the propagation factors removed). The modes are normalized such that $\langle \mathbf{Q}_n | \mathbf{Q}_m \rangle = \delta_{n,m}$. The equivalence between Eqs. (8) and (A1) was demonstrated in Ref. [27] assuming periodic lateral walls,

but it can be verified that the proof remains valid if other "closed" boundary conditions are enforced.

## B. Opaque-type boundaries

The equivalence between Eq. (A1) and the Chern number written as a function of the Berry curvature [Eq. (7)] is well-established when the boundary conditions are periodic [27, 34]. Next, I show that for other boundary conditions (e.g., PEC) the two formulas may give dissimilar results (see Eq. (11)).

As a starting point, it is observed that from Eq. (2) it follows that:

$$\partial_{\mathbf{k}} \hat{H}_g = \frac{1}{i}\left[\mathbf{r}, \hat{H}_g\right], \tag{A4}$$

where $\left[\hat{A}, \hat{B}\right] \equiv \hat{A}\hat{B} - \hat{B}\hat{A}$ stands for the commutator of two operators. Note that "$\partial_{\mathbf{k}} \hat{H}_g$" plays in photonics a role analogous to the "velocity operator" in quantum mechanics, and hence can be linked to the commutator of the position operator ($\mathbf{r}$) with the Hamiltonian. In the photonic case, the operator $\frac{1}{c^2}\partial_{\mathbf{k}}\hat{H}_g$ determines the Abraham (kinetic) light momentum in the cavity [41-43]. Indeed, it can be easily checked with the help of Eq. (A3) that

$$\left\langle \mathbf{Q} | \frac{1}{c^2}\partial_i \hat{H}_g | \mathbf{Q} \right\rangle = \frac{1}{c^2}\int_V dV \frac{1}{2}\hat{\mathbf{u}}_i \cdot \left(\mathbf{E} \times \mathbf{H}^* + \mathbf{E}^* \times \mathbf{H}\right), \tag{A5}$$

and the right-side is precisely the *i*-th component of the (Abraham) light-momentum of a (complex-valued) vector field [42, 43]. Here, $\mathbf{E}, \mathbf{H}$ are the electromagnetic field components of the state vector $\mathbf{Q}$.

Using Eq. (A4) and $\hat{H}_g \mathbf{Q}_n = \omega_n \mathbf{Q}_n$ it is seen that

$$\langle \mathbf{Q}_n | \partial_j \hat{H}_g | \mathbf{Q}_m \rangle = \frac{1}{i}(\omega_m - \omega_n)\langle \mathbf{Q}_n | x_j | \mathbf{Q}_m \rangle + R^{(j)}, \qquad (A6)$$

with $R^{(j)}$ defined as in Eq. (12) of the main text. Since $\hat{H}_g$ is an Hermitian operator it is tempting to set $R^{(j)} = 0$. However, here one needs to be rather careful because $\hat{H}_g$ is Hermitian *only* when the involved state vectors satisfy suitable boundary conditions. I will return to this issue shortly, first let us see the consequence of having $R^{(j)} = 0$. Substituting Eq. (A6) into Eq. (A2) it is readily found that:

$$\begin{aligned}\mathcal{F}_n &= \sum_{m \neq n} i\left[\langle \mathbf{Q}_n | x_1 | \mathbf{Q}_m \rangle \langle \mathbf{Q}_m | x_2 | \mathbf{Q}_n \rangle - 1 \leftrightarrow 2\right] \\ &= i\left[\langle \mathbf{Q}_n | x_1 x_2 | \mathbf{Q}_n \rangle - 1 \leftrightarrow 2\right] = 0\end{aligned} \qquad (A7)$$

In the second identity I used the completeness of the basis $\{\mathbf{Q}_m\}_{m=1,2,\ldots}$ and the fact the constraint $m \neq n$ can be dropped. The result $\mathcal{F}_n = 0$ is at first sight disconcerting because it implies that the Chern number calculated with Eq. (A1) [or equivalently with Eq. (8)] vanishes.

To bring some light into the discussion, it is observed that after some manipulations $R^{(j)}$ can be written explicitly as a surface integral over the cavity boundary ($\partial V$)

$$R^{(j)} = \frac{1}{2}\int_{\partial V} ds\, \hat{\mathbf{n}} \cdot \left[\mathbf{E}_n^* \times \mathbf{H}_m + \mathbf{E}_m \times \mathbf{H}_n^*\right] x_j, \qquad (A8)$$

where $\hat{\mathbf{n}}$ is the outward normal unit vector. Let us suppose first that the boundary conditions on the lateral walls are periodic. In this situation, the modes $\mathbf{Q}_n$ and $\mathbf{Q}_m$ are periodic, and thereby $\mathbf{E}_n^* \times \mathbf{H}_m + \mathbf{E}_m \times \mathbf{H}_n^*$ is also a periodic function. However, $\left[\mathbf{E}_n^* \times \mathbf{H}_m + \mathbf{E}_m \times \mathbf{H}_n^*\right] x_j$ is *not* periodic, and hence its flux over the lateral walls does not need to vanish, i.e., $R^{(j)} \neq 0$. In other words $\langle \hat{H}_g \mathbf{Q}_n | x_j | \mathbf{Q}_m \rangle \neq \langle \mathbf{Q}_n | \hat{H}_g x_j | \mathbf{Q}_m \rangle$, because

-34-

the function $x_j \mathbf{Q}_m$ does not satisfy periodic boundary conditions. The result $R^{(j)} \neq 0$ is reassuring because otherwise the Chern number ($\mathcal{C}_p$) calculated with Eq. (A1) would always vanish, which evidently cannot happen.

Let us consider now that PEC boundary conditions are imposed on the lateral walls. Clearly, in this case the state vector $x_j \mathbf{Q}_m$ satisfies the same boundary conditions as $\mathbf{Q}_m$, and thereby $R^{(j)} = 0$. As a by-product, it follows that the Chern number $\mathcal{C}_{\text{PEC}}$ calculated with Eq. (A1) [or with Eq. (8)] is indeed zero when the lateral walls are PEC, $\mathcal{C}_{\text{PEC}} = 0$, as I wanted to prove [Eq. (11)]. Clearly, the boundary conditions enforced on the Green function in Eq. (8) are of crucial importance.

As discussed in the main text, the lateral-walls are said to be "opaque" when the boundary conditions guarantee that $R^{(j)} = 0$. For opaque boundaries, one has $\mathcal{F}_n = 0$ [Eq. (A7)]. In electronics, the Berry curvature $\mathcal{F}_n$ can be understood as the normalized electric conductivity contribution from an electron in the *n*-th state when it is excited by a static-electric field [34]. In a system with periodic-type boundaries the electric current can be nonzero because the boundaries are *cyclic*, and hence it is possible to have $\mathcal{F}_n \neq 0$. In contrast, for opaque-type boundaries it is unfeasible to have a steady electric current because it cannot go through the boundaries, and therefore $\mathcal{F}_n = 0$. Translating these ideas to optics, one may regard $\mathcal{F}_n$ as the (linear) response of the light momentum due to some steady external stimulus (some analogue of the static-electric field excitation in fermionic systems), when the initial state of the system is determined by the *n*-th cavity mode. For cyclic (periodic) boundaries the induced momentum can be nontrivial, but for opaque-type boundaries it must vanish.

# Appendix B: Angular momentum expectation from the fluctuation-dissipation theorem

The fluctuation dissipation theorem establishes that the electromagnetic field correlations can be expressed in terms of the system Green-function (10) as [24, 40, 58]:

$$\frac{1}{(2\pi)^2}\left\langle\left\{\hat{\mathbf{f}}(\mathbf{r},\omega)\hat{\mathbf{f}}^\dagger(\mathbf{r}',\omega')\right\}\right\rangle_T = \delta(\omega-\omega')\mathcal{E}_{T,\omega}\frac{-1}{2\pi}\left[\overline{G}(\mathbf{r},\mathbf{r}',\omega)+\overline{G}^\dagger(\mathbf{r}',\mathbf{r},\omega)\right]_{\omega+0^+i}. \quad \text{(B1)}$$

Note that the definition of Green function used in this article [Eq. (10)] differs slightly from that of Ref. [24]. In the above, $\hat{\mathbf{f}} = \begin{pmatrix}\hat{\mathbf{E}} & \hat{\mathbf{H}}\end{pmatrix}^T$ represents the electromagnetic field quantum operator, $\{...\}$ is the symmetrized product of two operators, and $\langle ... \rangle_T$ gives the expectation at temperature $T$ [24]. Calculating the inverse Fourier transform of (B1) and using $\overline{\mathbf{G}}^*(\mathbf{r},\mathbf{r}',\omega) = \overline{\mathbf{G}}(\mathbf{r},\mathbf{r}',-\omega^*)$ one obtains the equal-time field correlations:

$$\left\langle\left\{\hat{\mathbf{f}}(\mathbf{r},t)\hat{\mathbf{f}}(\mathbf{r}',t)\right\}\right\rangle_T = -\frac{1}{\pi}\text{Re}\left\{\int_0^{+\infty}d\omega\,\mathcal{E}_{T,\omega}\left[\overline{G}(\mathbf{r},\mathbf{r}',\omega)+\overline{G}^T(\mathbf{r}',\mathbf{r},\omega)\right]_{\omega+0^+i}\right\}. \quad \text{(B2)}$$

In particular, the Poynting vector expectation $\langle\hat{\mathbf{S}}(\mathbf{r},t)\rangle_T = \left\langle\left\{\hat{\mathbf{E}}(\mathbf{r},t)\times\hat{\mathbf{H}}(\mathbf{r},t)\right\}\right\rangle_T$ may be written in a compact form as:

$$\langle\hat{S}_i(\mathbf{r},t)\rangle_T = -\frac{1}{\pi}\text{Re}\left\{\int_0^{+\infty}d\omega\,\mathcal{E}_{T,\omega}\,\text{tr}\left\{\overline{G}(\mathbf{r},\mathbf{r},\omega)\cdot\partial_i\hat{N}\right\}\Big|_{\omega+0^+i}\right\}, \quad \text{(B3)}$$

with $\partial_i\hat{N} = \begin{pmatrix}0 & -\hat{\mathbf{u}}_i\times\mathbf{1}_{3\times3} \\ \hat{\mathbf{u}}_i\times\mathbf{1}_{3\times3} & 0\end{pmatrix}$ and $i=1,2,3$. The expectation of the $z$-component of the angular momentum is thereby:

$$\langle\mathcal{L}_z\rangle_T = \int_0^{+\infty}d\omega\,\mathcal{L}_{T,\omega}, \quad \text{(B4a)}$$

$$\mathcal{L}_{T,\omega} = -\mathcal{E}_{T,\omega}\frac{1}{c^2}\frac{1}{\pi}\text{Re}\left\{\int dV\,\text{tr}\left\{\overline{G}(\mathbf{r},\mathbf{r},\omega)\cdot(x_1\partial_2\hat{N}-x_2\partial_1\hat{N})\right\}\Big|_{\omega+0^+i}\right\}. \quad \text{(B4b)}$$

In the above, $\mathcal{L}_{T,\omega}$ is the (unilateral) angular momentum spectral density at temperature *T*. The integration is over the cavity volume. The derived result is rather general and applies to lossy cavities (even in case of strong loss). The "quantum vacuum" expectation of $\mathcal{L}_z$ in a lossless cavity can be formally obtained from the fluctuation-dissipation theorem result simply by letting $T \to 0^+$ and by removing all the relevant dissipation mechanisms (e.g., by making the cavity walls perfectly conducting) [24]. This observation yields Eq. (21) where it is implicit that the Green function is evaluated for a lossless cavity.

## Appendix C: Partial angular momentum: fluctuation-dissipation theorem approach

In this Appendix, I derive Eq. (25) using as a starting point the fluctuation-dissipation theorem result (24). To begin with, I note that the integrand of Eq. (24) vanishes in the real frequency axis except at the locations of the cavity resonant frequencies. This is so because in the lossless-limit $\mathcal{L}_\omega$ consists of $\delta$-type functions centered at the resonant frequencies. Hence, integrating Eq. (24) by parts in frequency and supposing that $\omega_g$ is not coincident with a resonant frequency (for a bounded cavity the eigenfrequencies are countable and so this restriction is not problematic; the function $\mathcal{L}_z(\omega_g)$ is evidently discontinuous at the eigenfrequency values) one finds that:

$$\mathcal{L}_z(\omega_g) = \frac{1}{c^2}\frac{\hbar}{4\pi}\mathrm{Re}\int_{\omega_g}^{\omega_g+i\infty} d\omega\, \omega^2 \int dV\, \mathrm{tr}\left\{\partial_\omega \overline{\mathbf{G}}(\mathbf{r},\mathbf{r},\omega)\cdot\left(x_1\partial_2\hat{N}-x_2\partial_1\hat{N}\right)\right\}, \quad (C1)$$

with $\partial_\omega \equiv \partial/\partial\omega$.

Next, I note that from the definition of the Green function [Eq. (10)]

$$\left(\hat{N}-\omega\mathbf{M}(\mathbf{r},\omega)\right)\cdot\partial_\omega\overline{\mathbf{G}}(\mathbf{r},\mathbf{r}',\omega)=\partial_\omega\left[\omega\mathbf{M}(\mathbf{r},\omega)\right]\cdot\overline{\mathbf{G}}(\mathbf{r},\mathbf{r}',\omega). \tag{C2}$$

Furthermore, using $\hat{N}x_i = x_i\hat{N} - i\partial_i\hat{N}$ one obtains:

$$\left(\hat{N}-\omega\mathbf{M}(\mathbf{r},\omega)\right)\cdot x_i\partial_\omega\overline{\mathbf{G}}(\mathbf{r},\mathbf{r}',\omega)=x_i\partial_\omega\left[\omega\mathbf{M}(\mathbf{r},\omega)\right]\cdot\overline{\mathbf{G}}(\mathbf{r},\mathbf{r}',\omega)-i\partial_i\hat{N}\cdot\partial_\omega\overline{\mathbf{G}}(\mathbf{r},\mathbf{r}',\omega). \tag{C3}$$

A generic vector field $\mathbf{f}$ that satisfies $\left(\hat{N}-\omega\mathbf{M}(\mathbf{r},\omega)\right)\cdot\mathbf{f}(\mathbf{r})=i\mathbf{j}(\mathbf{r})$ and the same boundary conditions as the Green function has the integral representation:

$$\mathbf{f}(\mathbf{r}) = \int dV' \overline{\mathbf{G}}(\mathbf{r},\mathbf{r}',\omega)\cdot\mathbf{j}(\mathbf{r}'). \tag{C4}$$

Clearly, for opaque-type boundaries $x_i\partial_\omega\overline{\mathbf{G}}(\mathbf{r},\mathbf{r}',\omega)$ satisfies the same boundary conditions as the Green function. Thus, it has the integral representation:

$$x_i\partial_\omega\overline{\mathbf{G}}(\mathbf{r},\mathbf{r}',\omega)= \int dV''\overline{\mathbf{G}}(\mathbf{r},\mathbf{r}'',\omega)\cdot(-i)\left[x_i''\partial_\omega\left[\omega\mathbf{M}(\mathbf{r}'',\omega)\right]\cdot\overline{\mathbf{G}}(\mathbf{r}'',\mathbf{r}',\omega)-i\partial_i\hat{N}\cdot\partial_\omega\overline{\mathbf{G}}(\mathbf{r}'',\mathbf{r}',\omega)\right]. \tag{C5}$$

In particular, it follows that:

$$x_i\partial_\omega\overline{\mathbf{G}}(\mathbf{r},\mathbf{r},\omega)= \int dV'\overline{\mathbf{G}}(\mathbf{r},\mathbf{r}',\omega)\cdot\left[-ix_i'\partial_\omega\left[\omega\mathbf{M}(\mathbf{r}',\omega)\right]\cdot\overline{\mathbf{G}}(\mathbf{r}',\mathbf{r},\omega)-\partial_i\hat{N}\cdot\partial_\omega\overline{\mathbf{G}}(\mathbf{r}',\mathbf{r},\omega)\right]. \tag{C6}$$

This formula shows that the partial angular momentum [Eq. (C1)] can be decomposed as:

$$\mathcal{L}_z(\omega_g) = \frac{1}{c^2}\frac{\hbar}{4\pi}\mathrm{Re}\int_{\omega_g}^{\omega_g+i\infty}d\omega\,\omega^2\left[\mathcal{I}_1(\omega)+\mathcal{I}_2(\omega)\right], \tag{C7a}$$

$$\mathcal{I}_1 = -\int dV\int dV'\mathrm{tr}\left\{\partial_2\hat{N}\cdot\overline{\mathbf{G}}(\mathbf{r},\mathbf{r}',\omega)\cdot\partial_1\hat{N}\cdot\partial_\omega\overline{\mathbf{G}}(\mathbf{r}',\mathbf{r},\omega)-1\leftrightarrow 2\right\}, \tag{C7b}$$

$$\mathcal{I}_2 = \int dV\int dV'\mathrm{tr}\left\{(-ix_1')\partial_2\hat{N}\cdot\overline{\mathbf{G}}(\mathbf{r},\mathbf{r}',\omega)\cdot\partial_\omega\left[\omega\mathbf{M}(\mathbf{r}',\omega)\right]\cdot\overline{\mathbf{G}}(\mathbf{r}',\mathbf{r},\omega)-1\leftrightarrow 2\right\}. \tag{C7c}$$

The term associated with $\mathcal{I}_1$ reproduces Eq. (25) and hence it suffices to show that the term associated with $\mathcal{I}_2$ vanishes.

To do this, I use again Eq. (10) to find that $\left(\hat{N} - \omega \mathbf{M}(\mathbf{r},\omega)\right) \cdot x_i \overline{\mathbf{G}}(\mathbf{r},\mathbf{r}',\omega) = x_i i \mathbf{1} \delta(\mathbf{r}-\mathbf{r}') - i \partial_i \hat{N} \cdot \overline{\mathbf{G}}(\mathbf{r},\mathbf{r}',\omega)$. Hence, $\mathcal{I}_2$ can be expressed as:

$$\mathcal{I}_2 = \int dV \int dV'$$
$$\mathrm{tr}\left\{ x_1' \left[ \left(\hat{N} - \omega \mathbf{M}(\mathbf{r},\omega)\right) \cdot x_2 \overline{\mathbf{G}}(\mathbf{r},\mathbf{r}',\omega) - x_2 i \mathbf{1} \delta(\mathbf{r}-\mathbf{r}') \right] \cdot \partial_\omega \left[ \omega \mathbf{M}(\mathbf{r}',\omega) \right] \cdot \overline{\mathbf{G}}(\mathbf{r}',\mathbf{r},\omega) - 1 \leftrightarrow 2 \right\}$$
(C8)

After some simplifications and using the cyclic property of the trace ($\mathrm{tr}\{\mathbf{A} \cdot \mathbf{B}\} = \mathrm{tr}\{\mathbf{B} \cdot \mathbf{A}\}$), one can write:

$$\mathcal{I}_2 = \int dV \int dV' \mathrm{tr}\left\{ x_1' \partial_\omega \left[ \omega \mathbf{M}(\mathbf{r}',\omega) \right] \cdot \overline{\mathbf{G}}(\mathbf{r}',\mathbf{r},\omega) \cdot \left[ \left(\hat{N} - \omega \mathbf{M}(\mathbf{r},\omega)\right) \cdot x_2 \overline{\mathbf{G}}(\mathbf{r},\mathbf{r}',\omega) \right] - 1 \leftrightarrow 2 \right\}$$
(C9)

The integral representation (C4) is equivalent to $\mathbf{f}(\mathbf{r}') = \int dV \overline{\mathbf{G}}(\mathbf{r}',\mathbf{r},\omega) \cdot (-i) \left(\hat{N} - \omega \mathbf{M}(\mathbf{r},\omega)\right) \cdot \mathbf{f}(\mathbf{r})$, being implicit that the generic vector field $\mathbf{f}$ satisfies the same boundary conditions as the Green function. This result implies that:

$$x_2' \overline{\mathbf{G}}(\mathbf{r}',\mathbf{r}'',\omega) = \int dV \overline{\mathbf{G}}(\mathbf{r}',\mathbf{r},\omega) \cdot (-i) \left(\hat{N} - \omega \mathbf{M}(\mathbf{r},\omega)\right) \cdot x_2 \overline{\mathbf{G}}(\mathbf{r},\mathbf{r}'',\omega) \tag{C10}$$

Substituting this formula into Eq. (C9), one can see that:

$$\mathcal{I}_2 = \int dV' \mathrm{tr}\left\{ i x_1' x_2' \partial_\omega \left[ \omega \mathbf{M}(\mathbf{r}',\omega) \right] \cdot \overline{\mathbf{G}}(\mathbf{r}',\mathbf{r}',\omega) - 1 \leftrightarrow 2 \right\} = 0. \tag{C11}$$

This proves that the term associated with $\mathcal{I}_2$ really vanishes, and hence Eq. (25) follows.

## Appendix D: Partial angular momentum: a modal expansion approach

In this Appendix, I derive Eq. (25) relying on the modal-expansion (22), i.e.,

$$\mathcal{L}_z(\omega_g) = \sum_{\omega_n > \omega_g} \frac{\hbar \omega_n}{2} \mathcal{L}^{(n)}. \tag{D1}$$

The modes are normalized such that $\langle \mathbf{Q}_n | \mathbf{Q}_m \rangle = \delta_{n,m}$ and it is assumed that the cavity lateral walls are "opaque".

Using $\hat{H}_g = \sum_m \omega_m |\mathbf{Q}_m\rangle\langle\mathbf{Q}_m|$ and Eq. (A6) with $R^{(j)} = 0$ (for opaque-type lateral-walls) one may write $\mathcal{L}^{(n)}$ [Eq. (18)] as:

$$\mathcal{L}^{(n)} = \frac{1}{c^2} \sum_{m \neq n} \frac{\omega_m}{(\omega_m - \omega_n)^2} i \left[ \langle \mathbf{Q}_n | \partial_1 \hat{H}_g | \mathbf{Q}_m \rangle \langle \mathbf{Q}_m | \partial_2 \hat{H}_g | \mathbf{Q}_n \rangle - 1 \leftrightarrow 2 \right]. \tag{D2}$$

Substituting this formula into Eq. (D1) it follows that:

$$\mathcal{L}_z(\omega_g) = \frac{1}{c^2} \frac{\hbar}{2} \sum_{\substack{\omega_m < \omega_g, \\ \omega_n > \omega_g}} \frac{\omega_m \omega_n}{(\omega_m - \omega_n)^2} i \left[ \langle \mathbf{Q}_n | \partial_1 \hat{H}_g | \mathbf{Q}_m \rangle \langle \mathbf{Q}_m | \partial_2 \hat{H}_g | \mathbf{Q}_n \rangle - 1 \leftrightarrow 2 \right]. \tag{D3}$$

I used the fact that the generic term of summation is anti-symmetric with respect to interchanging the indices *m* and *n*. Due to this property the summation over *m* can be restricted to modes with $\omega_m < \omega_g$.

Similar to Appendix C, it is supposed that $\omega_g$ is not an eigenfrequency of the cavity. The line $\text{Re}\{\omega\} = \omega_g$ splits the complex-frequency plane into two semi-planes. It was shown in Ref. [27] that when $\omega_m$ and $\omega_n$ are in different semi-planes one has:

$$\int_{\omega_g - i\infty}^{\omega_g + i\infty} d\omega \frac{1}{(\omega - \omega_m)^2} \frac{1}{\omega - \omega_n} = \frac{2\pi i}{(\omega_m - \omega_n)^2} \text{sgn}(\omega_g - \omega_n). \tag{D4}$$

The same integral vanishes when $\omega_m$ and $\omega_n$ are in the same semi-plane. Hence, $\mathcal{L}_z(\omega_g)$ may be written as:

$$\mathcal{L}_z(\omega_g) = \frac{1}{c^2} \frac{-\hbar}{4\pi} \sum_{\substack{\omega_m < \omega_g, \\ \omega_n > \omega_g}} \int_{\omega_g - i\infty}^{\omega_g + i\infty} d\omega \frac{\omega_m}{(\omega - \omega_m)^2} \frac{\omega_n}{\omega - \omega_n} \left[ \langle \mathbf{Q}_n | \partial_1 \hat{H}_g | \mathbf{Q}_m \rangle \langle \mathbf{Q}_m | \partial_2 \hat{H}_g | \mathbf{Q}_n \rangle - 1 \leftrightarrow 2 \right]$$
(D5)

To proceed further, I note that by interchanging the indices $m$ and $n$ in Eq. (D3) one obtains $\mathcal{L}_z(\omega_g) = \frac{1}{c^2} \frac{-\hbar}{2} \sum_{\substack{\omega_n < \omega_g, \\ \omega_m > \omega_g}} [...]$ with the generic term of summation the same as in Eq. (D3). Hence, using again Eq. (D4) it is found that $\mathcal{L}_z(\omega_g) = \frac{1}{c^2} \frac{-\hbar}{4\pi} \sum_{\substack{\omega_n < \omega_g, \\ \omega_m > \omega_g}} \int_{\omega_g - i\infty}^{\omega_g + i\infty} d\omega [...]$ with the integrand the same as in Eq. (D5). Averaging this result and Eq. (D5) one finally obtains that:

$$\mathcal{L}_z(\omega_g) = \frac{1}{c^2} \frac{-\hbar}{8\pi} \sum_{m,n} \int_{\omega_g - i\infty}^{\omega_g + i\infty} d\omega \frac{\omega_m}{(\omega - \omega_m)^2} \frac{\omega_n}{\omega - \omega_n} \left[ \langle \mathbf{Q}_n | \partial_1 \hat{H}_g | \mathbf{Q}_m \rangle \langle \mathbf{Q}_m | \partial_2 \hat{H}_g | \mathbf{Q}_n \rangle - 1 \leftrightarrow 2 \right]$$
(D6)

The summation indices were unconstrained because the integral vanishes when $\omega_m$ and $\omega_n$ are in the same semi-plane. Noting that $\int_{\omega_g - i\infty}^{\omega_g + i\infty} d\omega \frac{1}{(\omega - \omega_m)^2} = 0$ one finds that:

$$\mathcal{L}_z(\omega_g) = \frac{1}{c^2} \frac{-\hbar}{8\pi} \sum_{m,n} \int_{\omega_g - i\infty}^{\omega_g + i\infty} d\omega \frac{\omega_m}{(\omega - \omega_m)^2} \frac{\omega}{\omega - \omega_n} \left[ \langle \mathbf{Q}_n | \partial_1 \hat{H}_g | \mathbf{Q}_m \rangle \langle \mathbf{Q}_m | \partial_2 \hat{H}_g | \mathbf{Q}_n \rangle - 1 \leftrightarrow 2 \right].$$
(D7)

The state vector can be decomposed as $\mathbf{Q}_n = \begin{pmatrix} \mathbf{f}_n & \mathbf{Q}_n^{(1)} & ... \end{pmatrix}^T$, with $\mathbf{f}_n$ the electromagnetic field associated with the $n$-th mode. Using the definition of the weighted inner product [Eq. (4)] and Eq. (A3) one sees that:

$$\mathcal{L}_z(\omega_g) = \frac{1}{c^2} \frac{-\hbar}{32\pi} \int_{\omega_g - i\infty}^{\omega_g + i\infty} d\omega \sum_{m,n} \frac{\omega_m}{(\omega - \omega_m)^2} \frac{\omega}{\omega - \omega_n} \iint dV dV' \left[ \mathbf{f}_n^*(\mathbf{r}) \cdot \partial_1 \hat{N} \cdot \mathbf{f}_m(\mathbf{r}) \mathbf{f}_m^*(\mathbf{r}') \cdot \partial_2 \hat{N} \cdot \mathbf{f}_n(\mathbf{r}') - 1 \leftrightarrow 2 \right]$$
(D8)

The photonic Green function has the modal expansion [27, 30, 59, 60]:

$$\overline{\mathbf{G}}(\mathbf{r},\mathbf{r}',\omega) = \frac{i}{2} \sum_n \frac{1}{\omega_n - \omega} \mathbf{f}_n(\mathbf{r}) \otimes \mathbf{f}_n^*(\mathbf{r}'). \tag{D9}$$

Hence, after some manipulations one obtains that:

$$\mathcal{L}_z(\omega_g) = \frac{1}{c^2} \frac{-\hbar}{8\pi} \int_{\omega_g - i\infty}^{\omega_g + i\infty} d\omega \iint dV dV' \left[ \mathrm{tr}\left\{ \omega \overline{\mathbf{G}}(\mathbf{r}',\mathbf{r},\omega) \cdot \partial_1 \hat{N} \cdot \partial_\omega \left[ \omega \overline{\mathbf{G}}(\mathbf{r},\mathbf{r}',\omega) \right] \cdot \partial_2 \hat{N} \right\} - 1 \leftrightarrow 2 \right] \tag{D10}$$

Integrating by parts in frequency the term "$1 \leftrightarrow 2$" and using the cyclic property of the trace, it is found that it is identical to the first term, so that:

$$\mathcal{L}_z(\omega_g) = \frac{1}{c^2} \frac{-\hbar}{4\pi} \int_{\omega_g - i\infty}^{\omega_g + i\infty} d\omega \iint dV dV' \, \mathrm{tr}\left\{ \partial_2 \hat{N} \cdot \omega \overline{\mathbf{G}}(\mathbf{r},\mathbf{r}',\omega) \cdot \partial_1 \hat{N} \cdot \partial_\omega \left[ \omega \overline{\mathbf{G}}(\mathbf{r}',\mathbf{r},\omega) \right] \right\}. \tag{D11}$$

The above formula gives $\mathcal{L}_z(\omega_g)$ as an integral over a straight-line parallel to the imaginary axis. Alternatively, it is possible to express $\mathcal{L}_z(\omega_g)$ as an integral over a path contained in the upper-half plane. To do this, I note that the term in rectangular brackets in Eq. (D8) is pure imaginary. Hence, a simple analysis shows that $\mathcal{L}_z(\omega_g) = \frac{1}{c^2} \frac{-\hbar}{32\pi} 2 \mathrm{Re} \int_{\omega_g}^{\omega_g + i\infty} d\omega [\ldots]$ with the integrand the same as in Eq. (D8) and "Re" the real-part operator. Using the modal expansion of the photonic Green function [Eq. (D9)] one obtains after some manipulations that:

$$\mathcal{L}_z(\omega_g) = \frac{1}{c^2} \frac{-\hbar}{4\pi} \mathrm{Re} \int_{\omega_g}^{\omega_g + i\infty} d\omega \iint dV dV' \left[ \mathrm{tr}\left\{ \partial_2 \hat{N} \cdot \omega \overline{\mathbf{G}}(\mathbf{r},\mathbf{r}',\omega) \cdot \partial_1 \hat{N} \cdot \partial_\omega \left[ \omega \overline{\mathbf{G}}(\mathbf{r}',\mathbf{r},\omega) \right] \right\} - 1 \leftrightarrow 2 \right] \tag{D12}$$

which gives the angular momentum as an integral along a straight line contained in the upper-half plane. The term $\partial_\omega[\omega\overline{\mathbf{G}}] = \overline{\mathbf{G}} + \omega\partial_\omega\overline{\mathbf{G}}$ originates two contributions, but the first one is invariant under a permutation of the indices "1" and "2". Thus, one obtains

$$\mathcal{L}_z(\omega_g) = \frac{1}{c^2}\frac{-\hbar}{4\pi}\mathrm{Re}\int_{\omega_g}^{\omega_g+i\infty}d\omega\,\omega^2\iint dVdV'\left[\mathrm{tr}\left\{\partial_2\hat{N}\cdot\overline{\mathbf{G}}(\mathbf{r},\mathbf{r}',\omega)\cdot\partial_1\hat{N}\cdot\partial_\omega\overline{\mathbf{G}}(\mathbf{r}',\mathbf{r},\omega)\right\} - 1\leftrightarrow 2\right]$$
(D13)

which yields the desired result [Eq. (25)].

Up to now, in this Appendix it was implicit that the cavity lateral-walls are "opaque" and hence the photonic Green function $\overline{\mathbf{G}}$ satisfies opaque-type boundary conditions. Let us now suppose that "periodic-type" boundary conditions are enforced on $\overline{\mathbf{G}}$. Furthermore, let $\omega_g$ lie in a band-gap of the corresponding photonic crystal. In these conditions, the integrand of Eq. (D11) is an analytic function in a vertical strip of the complex plane that includes the integration path (there are no poles in this strip). Hence, the integration path may be displaced in the strip without changing the value of $\mathcal{L}_z(\omega_g)$ calculated with Eq. (D11). Independent of the Green function boundary conditions, Eqs. (D11) and (D13) give the same result. Hence, it follows that $\mathcal{L}_z(\omega_g)$ calculated with Eq. (D13) with a Green function that satisfies periodic boundary conditions is a constant function, $\mathcal{L}_z = const.$, in each band-gap of the periodic structure. Using this result in Eq. (23), it follows that the spectral-density $\mathcal{L}_\omega$ vanishes in the band-gap [Eq. (26)] when $\overline{\mathbf{G}}$ satisfies periodic boundary conditions.